\documentstyle[aps,preprint,epsf,epsfig,rotate,prl]{revtex}
\tightenlines
\newcommand{\be}{\begin{equation}}
\newcommand{\ee}{\end{equation}}
\newcommand{\ba}{\begin{eqnarray}}
\newcommand{\ea}{\end{eqnarray}}

\draft        
\begin{document}             
\title{Absorption Line Shape of a One-Dimensional Bose Gas}
\author{S. K. Yip}
\address{Institute of Physics, Academia Sinica, Nankang, Taipei 11529, Taiwan}
%\date{\today}
\maketitle          

\begin{abstract}
We discuss the line shape for an  internal state transition  
for bosonic atoms confined as a one-dimensional gas.
Typical line shape is an edge singularity due to
the absence of Bose-Einstein Condensation
in such systems.

\end{abstract}

\vskip 0.2 cm

Recent progress in trapped bosonic gases has stimulated
a lot of activities in the study of many-body effects
in such systems.  Most theoretical papers concentrate
on effects due to the existence of the Bose-Einstein
Condensate (BEC) and the associated macroscopic wave-function.
With current techniques, it is likely that
one can as well study an effectively one-dimensional (1D)
trapped gas \cite{Olshanii98}.  
This situation occurs when the bosons are put in
a trap which is tightly confining in two directions
( $ y$ and $z$) but is essentially free in the third ($x$),
and with particles occupying only the lowest quantized
subband for motions in the $y$-$z$ plane.
This system is unique
in that, for any finite interaction among the bosons,
there is no Bose-Einstein Condensation even at zero temperature.
\cite{book}.  The long-wavelength quantum fluctuations
destroy the phase coherence of the system in the
macroscopic limit.  
For a given density $n$, the occupation number $N_0$ of the lowest
single particle state, though in general increases with
the size and hence the total number of particles $N$ in the system,
is nevertheless only a negligible fraction of the total 
($ {{\rm lim} \atop {N \to \infty}}  {N_0 \over N} \to 0$ ).
Though many properties of 1D bose gas have been studied in the
past \cite{book}, the subject remains somewhat academic
since there exists no physical realization.  
The possibility of realizing this 1D bose gas has
already stimulated some more theoretical work on
this system recently \cite{Rojo99,Girardeau00}

The present 1D bosonic system is closely related to its fermionic 
counterpart;
and they are collectively known as the ``Luttinger liquids''
\cite{Haldane81}.  One-dimensional fermionic systems have been
studied much more extensively in the context of
quantum wires, carbon nanotubes \cite{Tarucha95,Yao99} etc.
There the one-dimensional nature leads to
the absence of fermi liquid behavior, so
the properties of the system is qualitatively different
from its non-interacting counterpart.

The atomically trapped 1D bose gas offers us an excellent
new opportunity to study the peculiar properties of
quantum systems in reduced dimensions, in particular
the consequence due to the absence of BEC.
Not only that this is the first 1D bosonic system
available, it provides advantages over the mentioned
1D fermionic cases in that the present system
is intrinsically clean, and moreover numerous atomic
and optical tools can now be used.  It may
even be possible to further study novel systems
such as mixtures etc.

In this paper we consider one such example, namely
the absorption line shape of an optical transition
in this 1D bose gas. 
We imagine initially a quasi-1d system of bosons
 in their internal ground state
(referred to as the `a-atoms' below).
An incident optical wave excites one of the atoms
to a different internal state.  
We are interested in the absorption line shape
of this process; {\it i.e.}, the probability
that the absorption takes place at a given frequency $\Omega$.
 We shall consider the problem
at $T = 0$, paying particular attention to the
special feature due to the absence of BEC.  
An analogous
experiment has already been done
in a {\it three} dimensional trap 
%in the context of absorption line {\it shift} 
for the 1s $\to $ 2s
transition in hydrogen  \cite{Killian98}.
For a {\it uniform} 3D bose gas at $T=0$,
the line is expected to be narrow, Lorentzian
(see, e.g. \cite{Oktel99}, c.f. however \cite{Killian98})
with a width governed by the small `gaseous 
parameter(s)' $  (n_{3d} a^3 )^{1/2}  $.  Here
$n_{3d}$ is the particle number density in 3d
and $a$('s) is (are) the s-wave scattering length(s) for
scattering among the bosons.
For this three dimensional case, the main weight
of the transition comes from exciting a particle
from momentum $ \vec p = 0$.  There is a macroscopic 
number of such particles.  For a weakly interacting
gas, the potential felt by the rest of the bosons
is only slightly modified after the transition.
We shall see that the situation in 1D is very different
due to the absence of BEC. 
Exciting a single particle necessitates a substantial
rearrangement of the relative motion among
the rest of the particles, i.e., emission of a
large number of phonons. 
%even at $T = 0$
The basic line shape is typically an edge singularity.
%  This unique situation is due to
%the fact that a dilute 1d bose gas actually
%corresponds to where the interaction is {\it strong}.
%\cite{Lieb63}

We shall then consider an atomic trap with tight
confinement in the $y$-$z$ directions.  For definiteness
we shall assume that the confinement potentials in
these directions are isotropic and harmonic, with
frequency $\omega_{\perp}$.  We shall consider the
case where the atoms are limited within
$0 < x < L$ but otherwise free.
 A weak harmonic trap potential
along the x-axis has been considered in the past
\cite{Ho99,Petrov00}, and the low dimensional effects have been shown to
be significantly reduced.
We shall assume that such a potential is absent here.
We shall further ignore the effect due to finiteness of $L$,
a condition which we shall return below.

The bosonic system is described by a Hamiltonian containing
the kinetic energy and a short-range interaction among
the bosons.  Anticipating that we shall eventually study
the system  in the quasi-1d limit, the field operator
for the original bosons, referred to hereafter as the
a-bosons, is expanded as
%\be
$\psi_a(\vec r) =  \  \psi_a(x)
\chi_0 (\vec r_{\perp} ) \ 
+ \ 
\sum_j \ \psi_{aj}(x)
\chi_j (\vec r_{\perp} ) $
%\label{exp}
%\ee \noindent 
where $\psi_a(x)$ is the annihilation operator
for the a-bosons in the lowest subband ($0$) where
the transverse wavefunction is $ \chi_0 (\vec r_{\perp})$.
$\psi_{aj}(x)$'s ( $ j = 1, 2, ...$) are the corresponding
operators for the higher subbands (wavefunctions 
$ \chi_j (\vec r_{\perp})$).  In the ground state $ | G > $,
$\psi_{aj}(x) |G> = 0$ since all particles reside in the lowest
subband. 
We shall see that $\psi_{aj}$ for $j = 1, 2, ...$ will
not appear anywhere below, and can pretend that
the expansion of $\psi_a(\vec r)$ will consist thus
only the term involving the lowest subband.
%in eq (\ref{exp}).
The extent of the wavefunction $\chi_0$ in the y-z
direction will be denoted as $a_{\perp}$.  For
harmonic trapping potentials in these directions
$a_{\perp} = \sqrt { \hbar \over m \omega_{\perp }}$.

We shall not write down the interaction part of the Hamiltonian
involving the a-bosons explicitly since we won't need it below.
For delta-function type interactions it is possible to
write down all the formal many-body wavefunctions \cite{Lieb63}.
Unfortunately manipulations of such wavefunctions are typically
very mathematically involved.  Moreover we shall eventually
be interested in `final' states where an a-atom has been excited
internally. Such wavefunctions, where effectively
there is a `foreign' atom present, are not known
(with an exception noted below).  
We thus proceed rather differently and 
follow Haldane \cite{Haldane81}, concentrating
only on the low energy excitations which are density 
oscillations in the system.  The field operator 
$\psi_a(x)$  ( $ = \sum_p a_p  e^{i p x } / {\sqrt L}$) 
describing the motion along $x$ is re-expressed in terms of
density $n(x)$ and phase $\phi(x)$ by
$\psi_a(x) = [ n(x) ] ^{1/2} e^{i \phi(x)}$.  The effective
Hamiltonian for the 1d motion can be written as

\be
H_0 = {\hbar \over 2 \pi} \int dx 
\left[ \ vK \ ( \nabla \phi)^2  
 + 
{v \over K}
(\nabla \theta ) ^2 
\right]
\label{H0}
\ee

\noindent where $\nabla \theta$ is related to the number density
fluctuations $\delta n(x)$ by $\nabla \theta = \pi \delta n$.
$\phi(x)$ and $\theta(x)$ must be considered as operators
satisfying the commutation relation
$ [ \phi(x) , \nabla_{x'} \theta (x') ]  = i \pi \delta(x-x')$.
$v$ is the  density (sound) velocity wave of the system,
and $K$ is the (dimensionless) Luttinger liquid interaction parameter.
$ vK = \pi n_o \hbar / m$ and $K$ is related to the compressibility
of the system via  
$K = \pi v \hbar ({\partial n_o / \partial \mu} )$ where 
$n_o$ is the linear number density and
$\mu$ is the chemical potential of the system.
$K$ depends on the interaction among the bosons.
For short range interactions,
$K$ in principle have already been obtained \cite{Haldane81,Lieb63}
but there is no general analytic form known.  We simply note here that
$K = 1$ for the strongly repulsive (impenetrable) limit while $K \to \infty $ 
if the interaction among the bosons is weak.

$H_0$ can be diagonalized easily \cite{Haldane81} since it is quadratic.
It is useful to introduce bosonic operators  $b_q$ and $b_q^{\dagger}$
which describe the sound modes of the system:

\ba
\theta(x) &=&  
 - i \sum_{q \ne 0} | {\pi  K \over 2 q L} |^{1/2}
      sgn (q) e^{ i q x} ( b_{-q}^{\dagger}  + b_{q} ) 
\nonumber \\
\phi(x) &=&   
 - i \sum_{q \ne 0} | {\pi  \over 2 q L K } |^{1/2}
      e^{ i q x} ( b_{-q}^{\dagger} - b_{q} ) 
\label{transf}
\ea
\noindent Then, apart from a constant,  $H_0 = \sum_{q \ne 0} 
\hbar \omega_q b_q^{\dagger} b_q $ where the mode frequencies
are $\omega_q = v |q|$.

Now we consider the external optical field responsible
for the transition in internal state.  Since the
atom with a new internal state is distinguishable from
the original atoms, we shall call the resulting
atom the `c-atom'.  For definiteness we 
consider only the `Doppler-free' part of
the spectrum ( no momentum transfer from
the exciting laser beam(s)) and assume
that the perturbation responsible for the
 excitation is uniform throughout the 1D bose gas.
The relevant part of the Hamiltonian can be then written as
$H_{\rm ex} = w \int d^3 \vec r 
\left( \psi_c^{\dagger} (\vec r) \psi_a (\vec r)
+ 
 \psi_a^{\dagger} (\vec r) \psi_c (\vec r) \right)$
%\label{Hex} \noindent
where $\psi_c (\vec r)$ is the field operator
for the atom in the excited internal state.  We are interested
in the rate of transition as a function of
the excitation frequency $\Omega$.
This rate $I(\Omega)$ is given simply by the golden rule:
\be
I (\Omega ) = 2 \pi \sum_F | < F | H_{ex} | G > |^2 
\delta ( \Omega - ( E_F - E_G) )
\label{gold}
\ee
where $|F>$ are the set of final states.
%which is proportional to the imaginary part of
%the Fourier transform of 
%the retarded Green's function 
%\be
%{\sc D}^R (t) \equiv - i  < \hat X (t) \hat X^{\dagger} (0) > 
%h(t)
%\label{DR}
%\ee
%where $\hat X^{\dagger} \equiv \int d^3 \vec r  
%\psi_c^{\dagger} (\vec r) \psi_a (\vec r)$
%and the expectation value $ < ...> $ is 
%evaluated in the ground state $ | G > $ of the system of 
%$N$ a-atoms.  Here $ h(t)$ is the Heaviside step function.
%Since all a-atoms are in their lowest subband
%$\hat X^{\dagger}$ can be replaced by $ \int d^3 \vec r  
%\psi_c^{\dagger} (\vec r) \chi_0 (\vec r_{\perp}) \psi_a (x)$.
To proceed further we need to know the fate of the c-atom.
There are many possibilities and we shall simply consider
two extreme limits:

(1)  An atom in the internal state c is not affected at
all by the trapping potential for the atoms in state a.  This is
possible if, e.g., this potential is due to a
laser with frequency near a dipole resonance for the a-atoms
but far away from any of those for c.
In this case the c-atom escapes from the trap and no
longer interacts with the remaining a-atoms.

(2)  The c-atom also feels the strong transverse confinement potential.
It remains inside the trap and continues to interact with
the a-atoms.

We shall treat these two cases in turn.

{\it Case 1:}  \ 
In this case the final state is simply a product state, 
$ | F >  = | \Phi > \times | \vec p > $, with
the $N -1$ a-atoms in state $ | \Phi> $ 
(with x momentum $-p_x$) and the c-atom in the plane wave 
state $ | \vec p > $ with momentum $\vec p$. 
The rate for transition into states with given $\vec p$ is
\be
I_{\vec p} (\Omega) = 2 \pi | f_{\vec p} |^2 
\sum_{\Phi} | < \Phi | a_{px} | G > | ^2 
\delta ( \Omega - ( E_{\Phi} + {\vec p}^2 / 2 m + u - E_{G} ) )
\ee
where $f_{\vec p} \equiv \int d^2 \vec r_{\perp} 
e^{ - i \vec p_{\perp} \cdot \vec r_{\perp}}
 \chi_0 (\vec r_{\perp} )$
is a form factor, and $u$ is the difference
in internal energy between the a- and c- atoms.
  This rate is thus proportional to
the  spectral function for annihilation of a particle
in the 1D bose gas (hereafter $ p_x \to p$ )
\be
B(p,\omega') \equiv \sum_{\Phi} | < \Phi | a_p | G>|^2
\delta (\omega' - ( E_{\Phi} - E_{G,N-1}) )
\ee
with $\omega' = \Omega + E_G - E_{G,N-1} - 
{\vec p}^2 / 2 m - u $ and
where $E_{G,N-1}$ is the ground state energy of $N-1$ atoms
in the 1D trap.
$B(p,\omega')$ is given by $ - { 1 \over \pi} {\rm Im} G_1(p,\omega')$
 where 
\be
G_1(x,t) \equiv i < \psi_a^{\dagger}(0,0) \psi_a (x,t) > h(t) 
\label{G1}
\ee
The required Green's function can be evaluated 
by finding the long distance and time behavior
of $G_1(x,t)$.  Substituting   
eq (\ref{transf}) into eq(\ref{G1})  one finds 
$G_1 \sim { 1 \over  (x^2 + v^2 t^2 ) ^{1/4K} } h(t) $.  
The Fourier transform $G_1 (p, \omega')$ and hence
$B(p,\omega')$ for a given $p$ has the scaling form
$ \sim  h (\omega' - v | p|) / ( \omega'^2  - v^2 p^2 )^{1-\alpha}  $ 
%for $\omega' > v | p| $ 
with $\alpha = { 1 \over 4K}$.
Since $ 1 < K < \infty$ the line shape is of the 
form of an edge singularity.
For the three dimensional bose gas, the corresponding $B(\vec p, \omega')$
is proportional to a delta function at the Bogoliubov mode
frequency (with an additional small incoherent part).
The lack of a delta function in the present 
case reflects the absence of a condensate.
If $\vec p$ can be independently measured this experiment
would be analogous to angle resolved photoemission spectroscopy
employed extensively in recent studies of high T-c cuprates
\cite{Randeria97}.

{\it Case 2:} \  
In this case the final result depends on the 
interaction between the a and c atoms, which 
has the general form
$ g_{ac}
\int d \vec r \psi_{c}^{\dagger} (\vec r) \psi_{c} (\vec r) 
\psi_{a}^{\dagger} (\vec r) \psi_a (\vec r) $.
Here $g_{ac}$ is defined in 3D and is related to
the 3D scattering length $a_{ac}$ by 
$g_{ac} = 4 \pi \hbar a_{ac} / m$ .
We express $\psi_c (\vec r)$ 
also in subbands $ \psi_c (\vec r) = \sum_j \psi_{cj} (x) \chi_{cj} 
(\vec r_{\perp})$.  The final states in eq (\ref{gold})
 in general can involve terms that are off-diagonal
in the band index $j$ for the c-atom due to the
interaction between a and c.  Such processes correspond
to the possibility that the transverse motion of the c-atom
be modified due to the interaction with the a-atoms.
For a tight trap however, these contributions are small
in the dilute limit.  They involve the
parameter $ n_{o} g_{ac} / \pi a_{\perp}^2 \omega_{\perp}
\sim  a_{ac} / l $.  Here 
$l = 1/ n_{o}$ is the average 
interparticle spacing among the a-atoms.  Ignoring
thus these contributions,  we have
$I(\Omega) = \sum_j | f_j|^2 I_j(\Omega)$, {\it i.e.},
the transition line then becomes a superposition of
`lines' involving transition to final states where the c-atom is within
a given $j$-th subband, 
with a weight given by 
the form factor $ f_j = \int d \vec r_{\perp} [ \chi_0 (\vec r_{\perp})
\chi_{cj}^{*} (\vec r_{\perp}) ] $.
The shape of each of these lines is determined
by ${\rm Im} [{\sc D}^R_j (\Omega)]$ where
\be
{\sc D}^R_j (t) 
= -i \int dx_1  \int dx_2
<   \psi_a^{\dagger} (x_1, t) \psi_{cj} (x_1, t)
   \psi_{cj}^{\dagger} (x_2) \psi_a (x_2) >  h(t)
\label{Dj}
\ee
which we shall now compute.
The effective Hamiltonian is given by the sum of $H_0$ involving
only the a-atoms and $H_{cj}$ which describes the motion
of the c-atom within its $j$-th subband and its interaction
with the a-atoms.  $H_0$ has already been given in
eq(\ref{H0}) and the effective $H_{cj}$ is given by
$H_{cj}^{(1)} + H^{\rm int}_{cj}$ where 
\be
H_{cj}^{(1)} =  \int dx 
\left\{ {\hbar^2 \over 2 m} 
{\partial \psi_{cj}^{\dagger} (x) \over  \partial x}
{\partial \psi_{cj} (x) \over  \partial x}
+ ( \epsilon_{cj} + u ) \psi_{cj}^{\dagger}(x) \psi_{cj}(x)
\right\}
\ee
consists of the `one-body' (kinetic, trap and internal energy) 
terms and 
\be
H^{\rm int}_{cj} =
g_{ac,jj}  \int dx \psi_{cj}^{\dagger} (x) \psi_{cj} (x) 
\psi_{a}^{\dagger} (x) \psi_a (x)
\ee 
is due to the interaction between the a and c atoms. 
Here $\epsilon_{cj}$ is the subband energy for the
c-atom in its j-th subband and 
$g_{ac,jj} = g_{ac} \int d^3 \vec r  
| \chi_{cj} \chi_0(\vec r_{\perp})|^2$ depends on $j$.
For small $j$ it
 is of order $ g_{ac} / \pi a_{\perp}^2 $ if
the trap potentials for a and c atoms are similar.

It is convenient to express ${\sc D}_j^R(t)$ in the
`interaction' picture:
\be
{\sc D}^R_j (t) 
= -i \int dx_1  \int dx_2 
<   \psi_a^{\dagger} (x_1,t) \psi_{cj} (x_1,t)
\hat T e^{ - i \int_0^t  H^{\rm int}_{cj} (t') dt'}
   \psi_{cj}^{\dagger} (x_2) \psi_a (x_2) >  h(t)
\label{Dint}
\ee
where $\hat T$ is the time-ordering operator 
and the expectation value is to be calculated with
the Hamiltonion $H_0 + H^{(1)}_{cj}$.
One can rewrite eq(\ref{Dint}) in momentum
space and expand the exponent involving the interaction.
$H^{\rm int}_{cj}$ then produces `scattering' terms where 
the momentum of the c-atom is changed from one value to
the other.  The resulting calculation does not
seem to be tractable analytically in general.  The problem  however
can be simplified significantly if we assume
that $g_{ac,jj}$ is small (condition given below) and concentrate again
on the low frequency limit, i.e., just above the threshold
for transition into the $j$-th subband 
(given by $ \xi_{cj} = \epsilon_{cj} - \epsilon_0
 - \mu + u + g_{ac,jj} n_{1d}$ to lowest order in $g_{ac}$).
 In this case
transitions must be made to states where the momenta $p$
of the c-atom are small.
Provided that $ p << mv$, then the kinetic energy 
(of motion along x)  transferred to
the c-atom $ p^2 / 2m$ is much less than that to the
density oscillations of a-atoms $ v p $.  Thus so long 
as we are interested in frequency deviations $\Delta \Omega$ 
from the threshold which satisfy $\Delta \Omega << m v^2 $,
one can ignore the kinetic energy of the motion of the c-atoms
along $x$.  For an impenetrable bose gas this amounts to
limiting ourselves to $\Delta \Omega << \pi^2 \hbar^2 / m l^2 $.
Ignoring thus the first term in $H_{cj}^{(1)}$ and rewriting
the result back in real space, we obtain
\be
{\sc D}_j^R(t) = -i L e^{ -i \xi_{cj} t }
 <   \psi_a^{\dagger} (0,t) \
\hat T  e ^{ - i \int_0^t {\tilde H_{a, int}} (t') dt' }
\psi_{a} (0) >  h(t)
\label{DX}
\ee
where the effective interaction Hamiltonian
${\tilde H_{a, int} } $ acts only on the a-atoms and
is given by $ g_{ac,jj} \psi_a^{\dagger} (0) \psi_{a} (0)$.
This result can be understood physically as follows:
the external optical field produces an a $\to$ cj transition at $t=0$
and at a general location say $x_0$, annihilating an a-atom
while producing a c-atom there in the j-th subband.
Since the $c$ atom moves with velocity $p/m$ whereas the 
sound waves move with the speed $v$, at small $p$ the c-atom
is essentially stationary.  For $t > 0$,
due to the c-atom created, a delta-function 
interaction at $x_0$ acts on the a-atoms. We are interested
in the overlap between the initial state with an a-atom destroyed
at $x_0$ and
the final states with this extra interaction potential.
This overlap is independent of the location
of the transition $x_0$ which then has been set to $0$. 
The problem has thus become similar to that of X-ray absorption
in solid state physics.   There initially ( $ t < 0$) one
has an electron gas in its ground state.  At $t = 0$
an X-ray photon is absorbed which creates a charged nuclei
with an extra electron added to the electron gas,
which at $ t > 0$ feels an extra local potential due
to the charged nuclei.
Notice however there are slight differences.  Here 
for $t < 0$ we have an equilibrium 1D interacting bose gas and
at $t > 0$ there is one {\it less} boson than initially.

The correlation function in (\ref{DX}) and thus the line
shape can be found analogous to the X-ray problem
\cite{Doniach74}.  The expectation value needed can be
rewritten as $ < \psi^{\dagger}(0) e^{- i \tilde H t} \psi(0) >$
where $\tilde H = H_0 + \tilde H_{int}$.  
$H_0 = \sum \hbar \omega_q b_q^{\dagger} b_q$ 
and $\tilde H_{int} = g_{ac,jj} \delta n(0)$
 can both be written in terms of 
$b_q$ and $b_q^{\dagger}$.  The annihilation operator
$\psi \sim e^{ i \phi}$ acts like a displacement operator
for these bosons:  $ e^{i\phi} b_q e^{ - i \phi}
= b_q -  |{ \pi \over 2 q L K } |^{1/2} $.  
Thus the expectation value is the same as
$ < \hat T e^{ -i \int_0^t ( g_{ac,jj} - \pi v_N ) \delta n(0,t') dt'} >$
calculated with the Hamiltonian $H_0$.  
Here we have defined $v_N \equiv v/K$. 
The calculation can be
easily done since $H_0$ is quadratic in the boson operators
$b_q$ and $b_q^{\dagger}$.  The long time dependence
is given by, apart from a part oscillating sinusoidally with $t$
which contributes to the line shift,
${\sc D}_j^R (t) \sim 1/t^{\alpha'_j}$ and thus the line
shape $ \sim 1 / [\Delta \Omega] ^{ 1 - \alpha'_j}$ with
$\alpha'_j = { 1 \over 2 K} [ 1 - { g_{ac,jj} \over \pi v_N \hbar} ] ^2 $.
Notice that the line shape is in general different
for transition to different subbands of the `c-atom'.

It is instructive to compare this result with that of the X-ray problem
when the interaction between the nucleus after the X-ray
and the electrons is modelled by a delta-function 
interaction of strength $V$. 
There the long time behavior of the
corresponding correlation function is given by $1/ t^{\alpha_X}$
and the line shape is $ \sim 1 / [\Delta \Omega] ^{1 - \alpha_X}$
with $\alpha_X = [ 1 + N(0) V ] ^2$, 
where $N(0)$ is the density of states
at the fermi level for the electron gas \cite{Doniach74}.
%(assuming $N(0) V << 1$.)
Here, the interaction strength is replaced by $g_{ac,jj}$ and
the density of states is replaced by $ { \partial n \over \partial \mu}
= {  1 \over \pi v_N \hbar} $. 
In both cases the power law line shapes (for $\alpha_X \ne 0$ and
$\alpha'_j \ne 0$) are due to `orthogonal catastrophe',
that the overlap between the state just after the excitation
and the new ground state of the system vanishes.
   There are however two differences:
the $+$ sign in $\alpha_X$ is replaced by a $-$ sign since
now there is one less boson at $ t > 0$ rather than
one more electron in the X-ray absorption problem.  
 The decay of $D^R_j(t)$ in time is slower ($\alpha'_j$ is smaller)
if $g_{ac} > 0$:  near the location of the excitation,  the 
reduction in local density of the a-atoms is also what 
 an interaction with $g_{ac} > 0 $ prefers.
The factor $ 1/2K$ arises from the fact that we have a
one-dimensional interacting
bose gas rather than a three-dimensional fermi liquid. 

As the repulsive interaction $g_{ac,jj}$ between the a and c atoms
increases, $\alpha'_j$ decreases and the line sharpens up.
The above quantitive result for the exponents $\alpha'_j$ 
however holds only 
for small $g_{ac,jj}$ ( $<< \pi v_N \hbar $ ).
In the limit where interactions among {\it all} the bosons
are strongly repulsive, one reaches the impenetrable limit where
the bosons cannot pass each other.  In this
limit the statistics, i.e., the indistinguishability
among the a-bosons and the distinguishability between
the a- and c- bosons become irrelevant.   One
can check that the wavefunctions written down
in \cite{Girardeau60} for a system of
identical bosons only can be generalized to the present case
if we replace one of the bosons by the foreign c-atom.
It can be further verified that the absorption line
shape becomes delta functions 
(one for each subband cj) in this impenetrable limit.

The experimental observation of the line shape discussed
above should be feasible.  
The possibility of obtaining a quasi-1D bose gas
has already been discussed in ref \cite{Olshanii98}.
%The gas is quasi-one dimensional
%so long it is sufficiently dilute:
%$ n^{3d} g / \omega_{\perp} << 1 $  or equivalently
%$ a << l$.  Moreover, the gas also becomes essentially
%impenetrable in this limit, since $p a_{1d} \sim { a_{\perp}^2 
%\over l a}$ can be made sufficiently small provided the
%trap is sufficiently tight \cite{Olshanii98}.
The above line-shape is applicable so long as
$\Delta \Omega {< \atop \sim} { \pi^2 \hbar \over m l^2 } \equiv \omega_l$.
For hydrogen atoms and with $l \sim 1 \mu m$, $\omega_l 
\sim 100 {\rm k Hz} $, much larger than the present available experimental 
resolutions.  It is sufficient for the temperature $T$ to
be small compared with the line-width.  The condition 
$ T << \omega_l$ is satisfied for $ T < \mu K$, a temperature
readily achievable.  A finite length $L$ of the 
system will break the line into a set of discrete sub-lines
with separations of order $\pi^2 \hbar / m L^2$, but
the shape of the `line' can still be observable as long as
$ L >> l$. 

I thank Ite Yu for a useful correspondence.

\end{document}